\documentclass[aip,reprint]{revtex4-2}
\usepackage[utf8]{inputenc}
\usepackage[T1]{fontenc}
\usepackage{epsfig}
\usepackage{float}
\usepackage{mathptmx}
\usepackage{gensymb}
\usepackage{dcolumn}
\usepackage{bm}
\usepackage{graphicx}
\usepackage{times}
\usepackage{float}
\usepackage{amssymb}
\usepackage{amsmath}
\usepackage{amsfonts}
\usepackage{enumitem} 
\usepackage{afterpage}
\usepackage{epsfig}
\usepackage{setspace}
\usepackage[dvipsnames]{xcolor}
\usepackage{titlesec}
\usepackage{upgreek}
\usepackage{mathtools}
\usepackage[sort&compress]{natbib}

\setcitestyle{open={},close={},numbers,comma,super}

\usepackage[pagebackref=false]{hyperref}
\hypersetup{
    colorlinks  =   true,
    linkcolor   =   BlueViolet,
    citecolor   =   BlueViolet,
    filecolor   =   BlueViolet,
    urlcolor    =   BlueViolet}
\usepackage[capitalise,nameinlink]{cleveref}
\usepackage{titlesec}
\titleformat{\section}[hang]{\small\bfseries\sffamily}{\thesection.}{0.5em}{\MakeUppercase}
\titlespacing{\section}{0pc}{1.2pc}{0.3pc}
\titlespacing{\subsection}{0pc}{1pc}{0.2pc}

\makeatletter
\renewcommand*{\fnum@figure}{{\normalfont\bfseries \figurename~\thefigure}}
\renewcommand*{\@caption@fignum@sep}{\textbf{ : }}
\makeatother

\begin{document}
\title{Design of micron-long superconducting nanowire perfect absorber \\ for efficient high speed single-photon detection}

\author{Risheng Cheng}
\affiliation{Department of Electrical Engineering, Yale University, New Haven, CT 06511, USA}

\author{Sihao Wang}
\affiliation{Department of Electrical Engineering, Yale University, New Haven, CT 06511, USA}

\author{Chang-Ling Zou}
\affiliation{Department of Electrical Engineering, Yale University, New Haven, CT 06511, USA}
\affiliation{Department of Optics, University of Science and Technology of China, Hefei 230026, Anhui, China}

\author{Hong X. Tang}
\email{hong.tang@yale.edu}
\affiliation{Department of Electrical Engineering, Yale University, New Haven, CT 06511, USA}

\date{\today}

\begin{abstract}
Despite very efficient superconducting nanowire single-photon detectors (SNSPDs) reported recently, combining their other performance advantages such as high speed and ultra-low timing jitter in a single device still remains challenging. In this work, we present a perfect absorber model and corresponding detector design based on a micron-long NbN nanowire integrated with a 2D-photonic crystal cavity of ultra-small mode volume,  which promises simultaneous achievement of near-unity absorption, gigahertz counting rates and broadband optical response with a 3\,dB bandwidth of 71\,nm. 
Compared to previous stand-alone meandered and waveguide-integrated SNSPDs, this perfect absorber design addresses the trade space in size, efficiency, speed and bandwidth for realizing large on-chip single-photon detector arrays.     
\end{abstract}
\maketitle

\section{Introduction}

Superconducting nanowire single-photon detectors (SNSPDs)\cite{goltsman_2001_first_SNSPD,Hadfield_2009_SPD_review} are recognized as one of the most important photon detection technologies in quantum information processing and communications\cite{yamamoto_2007_quantum_key_snspd,wang_2018_multidimensional_quantum,liao_2017_satellite}, due to their high internal quantum efficiency over a broad wavelength band\cite{nist_2013_93p_efficiency,simit_2017_92p_nbn_detector,delft_2017_92p_nbn_detector,nist_2019_mid-ir_detector_spectroscopy,berggren_2012_mid-ir_detector,cheng_2019_broadband_spectrometer}, fast speed\cite{pernice_2018_2DPC,simit_2019_16_pixel_detector}, excellent timing performance\cite{jpl_2018_low_jitter,Delft_2018_10ps_jitter_detector,JPL_2019_MoSi_detector_jitter} and ultra-low dark count rates\cite{schuck_2013_mHz_dark_count,NTT_2015_ultimate_darkcounts}. However, these individual performance characteristics have been so far best optimized in separate device designs, and it still remains a challenge to incorporate all the high performance merits in a single device, due to the inherent trade-off between detection efficiency and bandwidth over a desired nanowire length. For example, in most high-efficiency SNSPD designs, it is preferable that the devices consist of an ultra-thin and millimeter-long superconducting nanowire meandered into a circular shape with >$15\,\upmu\mathrm{m}$ diameter to guarantee near-unity coupling efficiency from fiber to detector\cite{nist_2013_93p_efficiency,simit_2017_92p_nbn_detector,delft_2017_92p_nbn_detector,cheng_2017_multiple_SNAP}, which renders the detectors subject to speed limited by the large kinetic inductance of the long nanowires\cite{berggren_2006_kinetic_inductance}. Integrating the nanowire detectors with on-chip optical waveguides could greatly increase the interaction time of the nanowire  with photons and thereby reduce the absorption length of the nanowire down to tens of micrometers\cite{Fiore_2011_waveguide_snspd,pernice_2012_waveguide_SNSPD,schuck_2013_nbtin_detector_sin_waveguide,pernice_2013_absorption_engineering,Fiore_2015_GaAs_SNSPD,pernice_2015_waveguide_snspd,pernice_2015_snspd_diamond,berggren_2015_on_chip_detector,hadfield_2016_MoSi_waveguide_detector,pernice_2018_waveguide_snspd_review,italy_2019_amplitude_multiplex}. Nevertheless, the total length of the nanowire needs further scale down below $10\,\upmu\mathrm{m}$ in order to realize gigahertz counting rates\cite{pernice_2016_1D_PhC_detector}. Moreover, the SNSPD with a record-low timing jitter below $3\,\mathrm{ps}$ is only demonstrated with a $5\,\upmu\mathrm{m}$-long NbN nanowire recently \cite{jpl_2018_low_jitter} in order to suppress the geometry-induced jitter\cite{berggren_2016_geometric_jitter}, albeit with a compromised detection efficiency due to direct fiber illumination.  Considerable efforts have been devoted to further reducing the nanowire length by embedding the nanowires into a variety of cavities, such as 1D-, 2D-photonic crystal (PhC) cavities\cite{canada_2015_perfect_nanowire_absorber,pernice_2016_1D_PhC_detector,pernice_2018_2DPC} and racetrack ring resonators\cite{Bristol_2016_modelling_snspd_wg_cavity}. While as short as $1\,\upmu\mathrm{m}$-long nanowires are demonstrated with a considerable absorption, the use of resonant cavities introduces spectral selectivity, resulting in an operation bandwidth of only several nanometers.  

The main goal of this work is to design a near-perfect absorber for an ultra-short nanowire of a micron length while simultaneously maintaining a broad operation bandwidth around the telecommunication wavelength of 1550\,nm. We begin by theoretically studying a general one-side cavity model in \cref{section:lossy} to guide the optimization of the nanowire loaded cavity system. Then, we systematically investigate the absorption rates of the waveguide-integrated NbN nanowires in dependence on the index contrast of waveguides, nanowire geometry and varying waveguide types to maximize the photon loss per round-tip in the cavity (\cref{section:absorption}).  In \cref{section:2D}, we present the design and simulation results of the ultra-short NbN nanowire integrated with an H0-type 2D-PhC cavity, which demonstrates an ultra-small mode volume and thus enables the nanowire to achieve a near-unity peak absorption efficiency combined with more than $70\,\mathrm{nm}$ 3\,dB bandwidth maintained. In \cref{section:comparison}, we compare our design with previous implementations of SNSPDs based on different device structures, and the results demonstrate orders of magnitude improvement achieved with our design in terms of the bandwidth-nanowire-length ratio, which holds promise for realizing a large array of high-performance on-chip single-photon detectors for future integrated quantum photonic circuits.    

\section{One-side cavity model}
\label{section:lossy}

\begin{figure}[!htb]
\centering\includegraphics[width=1\linewidth]{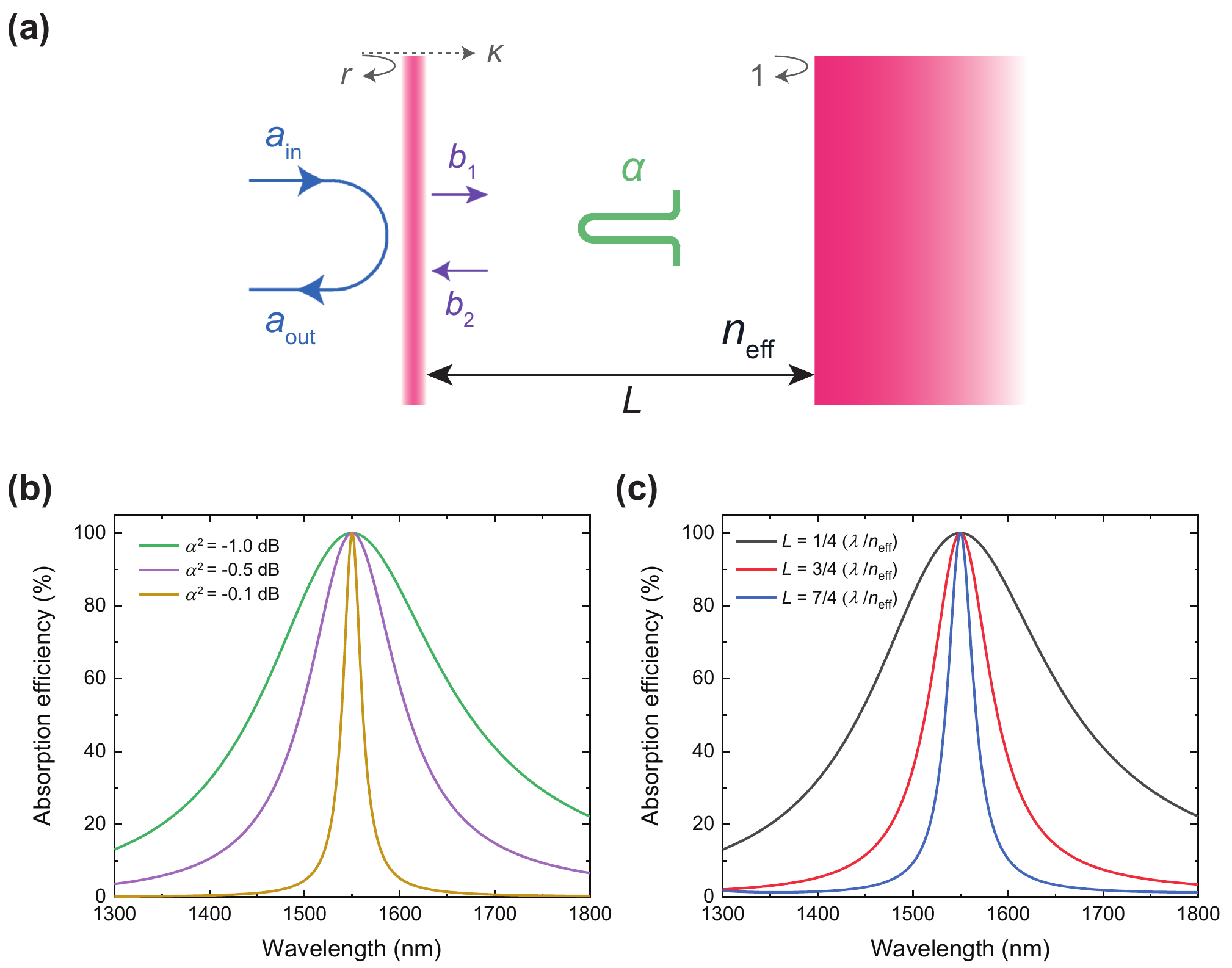}
\caption
{One-side cavity model and calculated absorption efficiency of the nanowire absorber.
(a) Schematic of the lossy cavity with an embedded nanowire absorber.
(b) Calculated nanowire absorption spectrum for varying photon loss per round-trip in the cavity. The cavity length is fixed at $L = {1}/{4} ({\lambda}/{n_\mathrm{eff}})$. 
(c) Calculated nanowire absorption spectrum for varying cavity length. The photon loss per round-trip is fixed at $\alpha^2=-1.0\,\mathrm{dB}$.
}
\label{fig:perfect_lossy}
\end{figure}

In order to study the effect of the cavity on the nanowire absorption bandwidth, we propose a simple and general one-side cavity model as illustrated in \cref{fig:perfect_lossy}(a). The nanowire is placed inside a cavity which is formed by a full reflective mirror combined with another partial reflective mirror. Through the partial mirror, light could couple into and out of the cavity with the designed coupling coefficient $\kappa$, while $r$ denotes the field amplitude reflection coefficient of the partial mirror.  We assume there is no loss in the coupling region, and thus the relation $r^2+\kappa^2 = 1$ holds. Using the transfer matrix method, we could relate the fields at the left and right side of the partial mirror as
\renewcommand{\arraystretch}{0.8}
\begin{equation}
\begin{bmatrix} 
a_\mathrm{in} \\ a _\mathrm{out}
\end{bmatrix}
=  \frac{1}{\kappa} 
\begin{bmatrix} 
1 & r \\ r & 1 
\end{bmatrix}
\begin{bmatrix} 
b_1 \\ b_2 
\end{bmatrix}
\label{eq:tmm}
\end{equation}

\vspace{0.5mm}
\noindent where $a_\mathrm{in}$ and $a_\mathrm{out}$ represent the incident field and the reflected field from the cavity, while $b_1$ and $b_2$ denote the field inside the cavity propagating forward and backward, respectively. $b_1$ and $b_2$ are further related by 
\begin{equation}
b_2 = b_1\alpha e^{-i\theta}
\label{eq:b12}
\end{equation}
\noindent and
\begin{equation}
\theta = 4\pi Ln_\mathrm{eff}/\lambda + \pi
\label{eq:phase_shift}
\end{equation}
\noindent where $\alpha$, $\theta$, $L$, $n_\mathrm{eff}$ and $\lambda$ represent the amplitude attenuation coefficient, the round-trip phase shift, the cavity length, the effective index of the cavity and the wavelength of the incident light, respectively. By substituting \cref{eq:b12} and \cref{eq:phase_shift} into \cref{eq:tmm}, we obtain the power absorption efficiency $A$ and the reflectance $R$ of the cavity as follows:   
\begin{equation}
\begin{split}
A &= 1-R \\
&= 1-\left|\frac{a_\mathrm{out}}{a_\mathrm{in}}\right|^2 \\
&= 1-\left|\frac{r+\alpha e^{-i\theta}}{1+r\alpha e^{-i\theta}}\right|^2 \\
&= 1- \frac{r^2+\alpha^2-2r\alpha\mathrm{cos}\theta}{1+(r\alpha)^2-2r\alpha\mathrm{cos}\theta}
\end{split}
\label{eq:absorption}
\end{equation}

In order to achieve a perfect absorption $A=1$, \cref{eq:absorption} indicates that two conditions must be fulfilled. (1) The cavity needs to be on resonance, i.e., $\lambda = \lambda_\mathrm{res} = 4n_\mathrm{eff}L/m$ or $L = m/4(\lambda_\mathrm{res}/n_\mathrm{eff})$, where $m$ is the longitudinal mode order number and has to be positive integers. (2)  The critical coupling condition\cite{yariv_2002_critical_coupling} should be satisfied, i.e.,$r = \alpha$ for odd $m$ and $r = -\alpha$ for even $m$. The negative value of $r$ indicates half-wave loss induced at the interface between the input medium and the cavity. This condition also leads to $1-\alpha^2 = \kappa^2$, suggesting that the coupled power into the cavity is equal to the loss. We note that $\alpha^2$ mentioned here stands for the total loss of the photons per round-trip in the cavity, including the desired absorption by the nanowire as well as the intrinsic loss in the cavity due to dielectric loss, scattering and radiation. Therefore, the designed cavity should have negligible intrinsic loss compared to the nanowire absorption to guarantee all the photons coupled into the cavity is finally absorbed by the nanowire.

From \cref{eq:absorption}, we could derive the loaded quality factor $Q_\mathrm{l}$ and the absorption bandwidth defined as the full width at half maximum (FWHM) of the resonance spectrum: 
\begin{equation}
\begin{split}
Q_\mathrm{l} 
&= \frac{\lambda_\mathrm{res}}{\mathrm{FWHM}} \\
&= \frac{2\pi n_\mathrm{eff}L\sqrt{|r|\alpha}}{\lambda_\mathrm{res}(1-|r|\alpha)}\\
&= \frac{m\pi \sqrt{|r|\alpha}}{2(1-|r|\alpha)}
\end{split}
\label{eq:loaded}
\end{equation}

\noindent which can be further simplified at the condition of critical coupling ($|r|=\alpha$) as
\begin{equation}
\begin{split}
Q_\mathrm{l,c} 
&= \frac{m\pi\alpha}{2(1-\alpha^2)}
\end{split}
\label{eq:loaded_resonance}
\end{equation}

\Cref{eq:loaded_resonance} suggests that in order to realize a broadband absorber with large FWHM, $Q_\mathrm{l}$ should be minimized, which can be achieved by simultaneously  employing a cavity of small mode volume (smaller $m$ or $L$) and enhancing the absorption efficiency of photons per round-trip in the cavity (smaller $\alpha$). These results are further quantitatively visualized by plotting the calculated nanowire absorption spectrum for varying $\alpha$ and $L$ in \cref{fig:perfect_lossy}(b) and (c), respectively. We ignore any intrinsic loss in the cavity and assume all the photon loss is due to the absorption by the nanowire. In \cref{fig:perfect_lossy}(b), we observe a significant absorption bandwidth (>\,200\,nm) with the nanowire absorption rate $\alpha^2 = -1\,\mathrm{dB}$, which can be easily obtained by NbN nanowires atop optimized high-index waveguides at micron length scale, as will be shown in the following \cref{section:absorption}. \Cref{fig:perfect_lossy}(c) demonstrates that the FWHM decreases sharply as the mode number $m$ increases, and thus the fundamental mode ($m=1$) is desired for the largest bandwidth. This can explain the broadband nature of the meander-type SNSPDs with vertically integrated quarter-$\lambda$ cavities\cite{nist_2013_93p_efficiency,simit_2017_92p_nbn_detector,delft_2017_92p_nbn_detector,simit_2016_broadband_snspd}, despite the minimum length of the nanowire is limited in another dimension. However, for on-chip integrated cavities, it is challenging to design a cavity with an extremely small mode volume combining low intrinsic loss due to the strong radiation and scattering of photons at optical frequencies. For example, to maintain an acceptable radiation loss due to waveguide bending, $m$ is typically designed larger than 20 for micro-ring resonators fabricated on the SOI platform \cite{Bogaerts_2011_ring_resonator_review} and even larger for other medium-index waveguide platforms\cite{guo_2017_photon_pair,fan_2018_eo_converter}. Fortunately, the advent of PhC nanocavities\cite{Yablonovitch_1987_first_photonic_crystal,Akahane_2003_high-Q_2D_phc_cavity} provides us the possibility of trapping light in a smaller-than-wavelength mode volume by inhibiting radiation via photonic bandgap effect, which will be further discussed in \cref{section:2D}.

\section{Study of nanowire absorption rates}
\label{section:absorption}

\subsection{High-index-contrast and  medium-index-contrast waveguides}

\begin{figure}[!b]
\centering
\includegraphics[width=\linewidth]{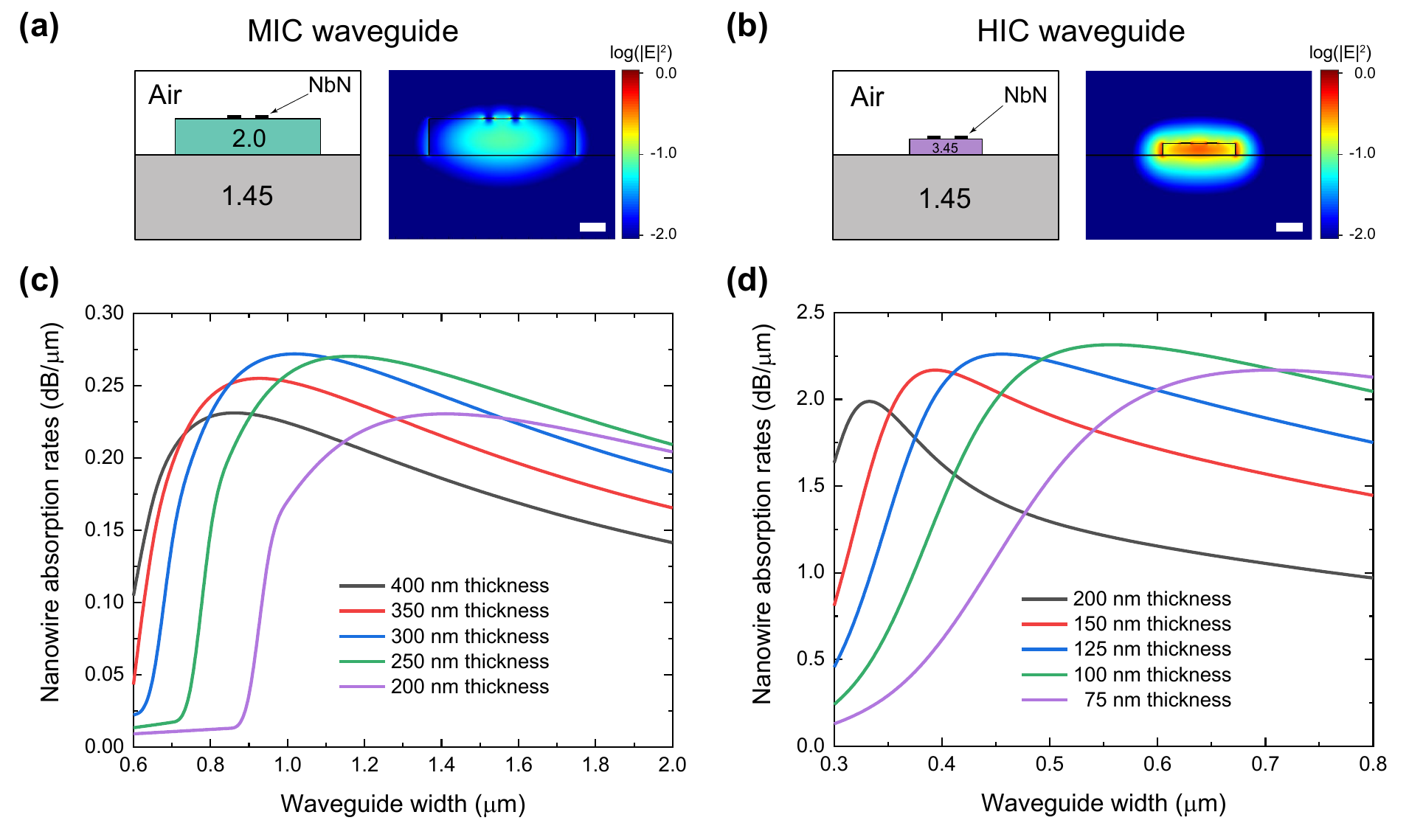}
\caption
{Comparison between HIC and MIC waveguides.
(a,b) Cross-sectional schematics and simulated mode profiles of (a) MIC and (b) HIC waveguides. Scale bars, 200\,nm. 
(c,d) Simulated nanowire absorption rates as a function of waveguide width for (c) MIC and (d) HIC waveguides of varying thickness.}
\label{fig:perfect_comparison}
\end{figure}

In this subsection, we first compare the absorption rates of NbN nanowires integrated with waveguides of different index contrast. As shown in the schematic drawings of \cref{fig:perfect_comparison}(a) and (b), we set the index of the core material as 3.45 and 2.0 for high- and medium-index-contrast (HIC and MIC) waveguides, respectively. The waveguides have the bottom cladding of 1.45 index and air upper cladding. The MIC waveguide could approximately represent a wide range of waveguide platforms, such as SiN\cite{schuck_2013_nbtin_detector_sin_waveguide,pernice_2015_waveguide_snspd}, GaN\cite{xiong_2011_integrated_gan}, AlN\cite{xiong_2012_aln_review,guo_2017_photon_pair}, diamond\cite{pernice_2015_snspd_diamond} and LiNbO\textsubscript{3}\cite{loncar_2018_integrated_linbo3,sayem_2019_ln_snspd} on insulators (SiO\textsubscript{2}), while GaAs\cite{bower_2018_gaas_on_insulator} and Si\cite{pernice_2012_waveguide_SNSPD} on insulators could be good examples of HIC waveguide. The U-shaped NbN nanowire is placed atop these waveguides both as absorber and detector via evanescent coupling to the optical mode confined in the waveguides. The complex refractive index of NbN we use in the simulation is $5.23 + i5.82$ measured by ellipsometry\cite{berggren_2008_snspd_optical_modeling}. The thickness, width and gap of the nanowire are set to 5$\,$nm, 80$\,$nm and 120$\,$nm, respectively, based on previous experimental results \cite{delft_2017_92p_nbn_detector,simit_2017_92p_nbn_detector,cheng_2019_snspd_ald}, which could guarantee saturated internal efficiency at 1550$\,$nm wavelength.

By numerically solving the fundamental eigenmode of the waveguide, we extract the effective index of the propagating mode, the imaginary part of which corresponds to the absorption. The nanowire absorption rates can be obtained using the expression $AR = 4.34(4\pi k_\mathrm{eff}/\lambda)$ in units of $\mathrm{dB}/\upmu\mathrm{m}$, where $k_\mathrm{eff}$ is the imaginary part of the effective index. 
We only consider fundamental TE modes in this subsection, since TE modes are more widely used in the PhC cavity design due to their larger photonic bandgap than TM modes.  

\Cref{fig:perfect_comparison}(c) and (d) plot the simulated results of the nanowire absorption rates as a function of the waveguide width for varying MIC and HIC waveguide thickness. For both waveguide designs, the absorption rates show smooth dependence on the waveguide width, and the maximum peaks shift towards smaller width for increased waveguide thickness. For wider and thicker waveguides, the absorption rates decrease due to better confined optical mode and thus weaker evanescent coupling to the nanowire. On the other hand, the absorption rates also drop for narrower and thinner waveguides because of increased mode size and  cut-off condition. As a result, the maximum absorption rate of 0.27$\,$dB/$\upmu$m is achieved with a 300$\,$nm-thick and 1$\,\upmu$m-wide MIC waveguide. For HIC waveguides, we observe almost one order of magnitude enhanced value of  2.3$\,$dB/$\upmu$m occurring around 100$\,$nm thickness and 550$\,$nm width. This huge difference is not surprising, considering more compact mode profile achieved with HIC waveguides in comparison with MIC waveguides, as shown in \cref{fig:perfect_comparison}(a) and (b). The simulation results are also in good agreement with previously reported experimental values\cite{pernice_2012_waveguide_SNSPD,schuck_2013_nbtin_detector_sin_waveguide,pernice_2013_absorption_engineering,pernice_2015_waveguide_snspd}. Therefore, we only focus on the optimization of HIC waveguide design in the following subsections.

\subsection{Effect of nanowire geometry}
\label{section:nw_geometry}

\begin{figure}[!b]
\centering\includegraphics[width=0.75\linewidth]{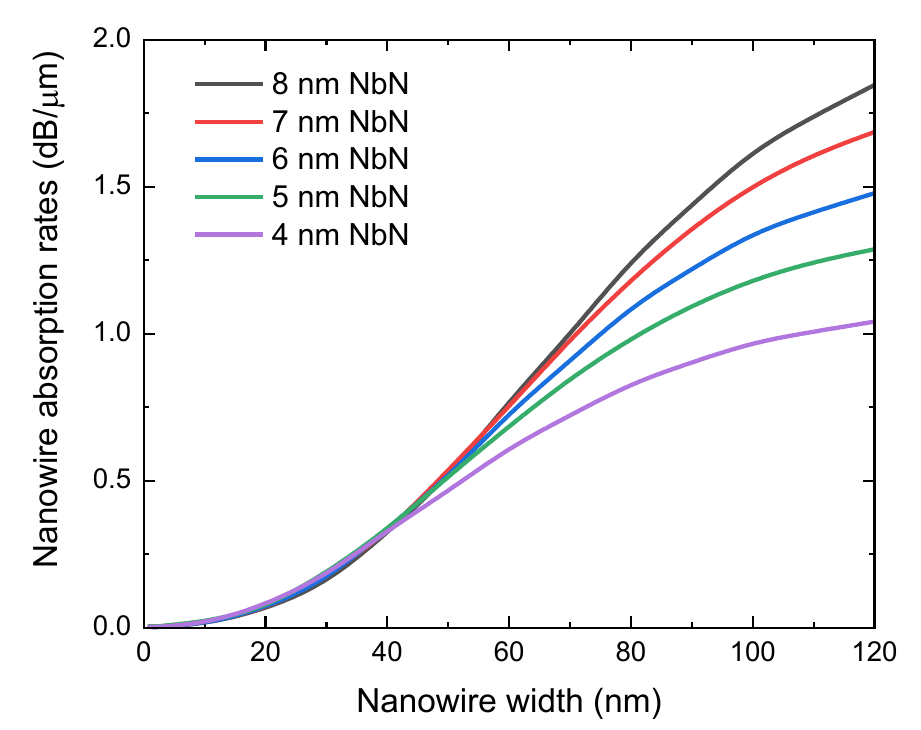}
\caption
{
Simulated nanowire absorption rates depending on the nanowire width and thickness for the HIC waveguide design. 
}
\label{fig:perfect_nw_geometry}
\end{figure}

\begin{figure*}[!t]
\centering\includegraphics[width=0.8\linewidth]{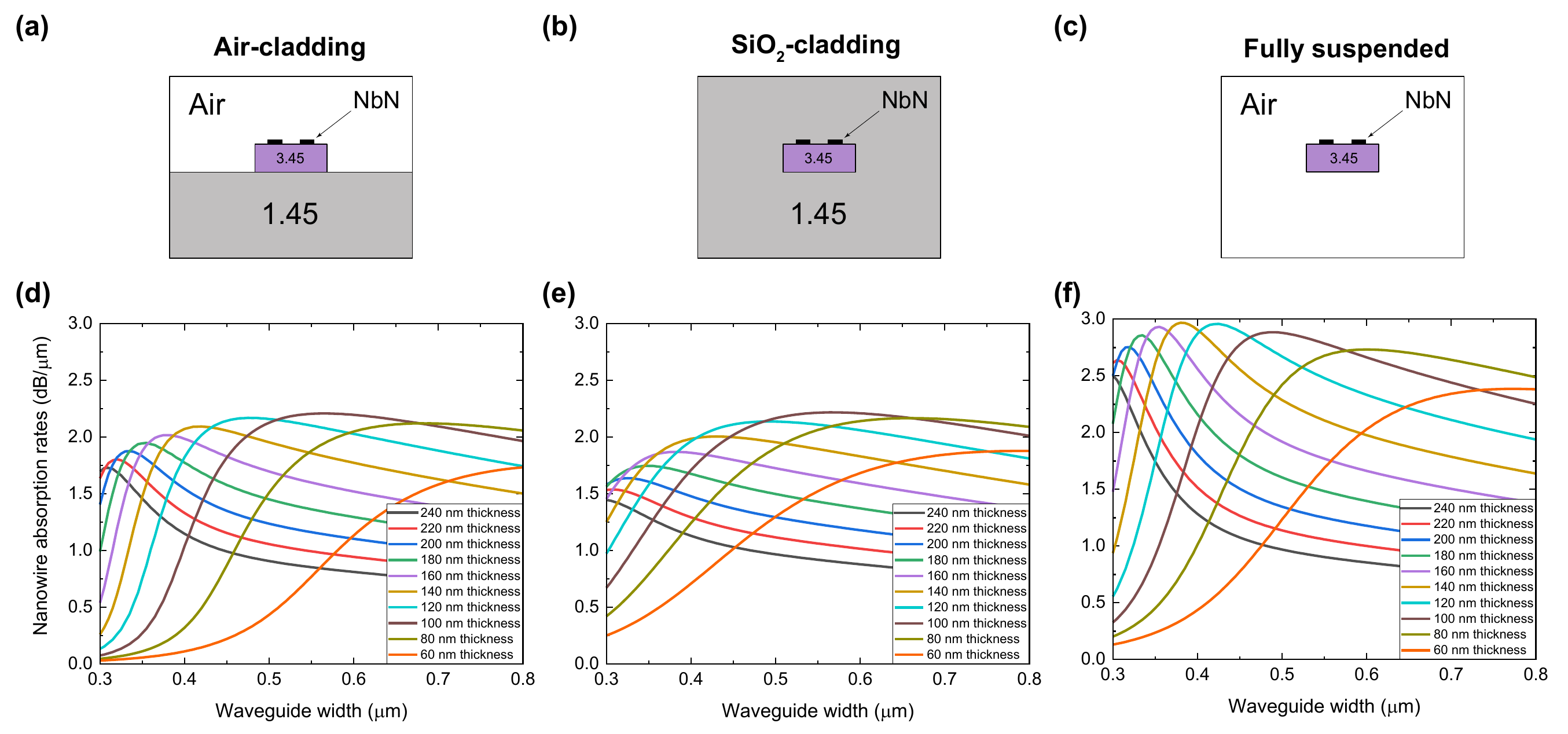}
\caption
{Effect of the waveguide types and geometry on the nanowire absorption rates.
(a-c) Cross-sectional schematics of (a) air-cladding waveguide on SiO\textsubscript{2}, (b) SiO\textsubscript{2}-cladding buried waveguide and (c) fully suspended waveguide. 
(d-i) Simulated nanowire absorption rates depending on the waveguide geometry for HIC waveguides with various upper and bottom claddings of different indices.
The corresponding simulation results are shown in the same column as the schematics of the waveguides.  
}
\label{fig:perfect_wg_type}
\end{figure*}

In \cref{fig:perfect_nw_geometry}, we show the effect of the nanowire geometry on its absorption rates. We use the HIC waveguide design shown in \cref{fig:perfect_comparison}(b) in the simulation and set the thickness and width to 220$\,$nm and 600$\,$nm. The gap between the nanowires are kept 1.5 times the width of the nanowire to relieve the current crowding effect\cite{berggren_2011_current_crowding}. Interestingly, we observe that the absorption rates are nearly independent of the nanowire thickness in ultra-narrow (<40$\,$nm width) region and start to grow rapidly for larger width beyond this region. We attribute this behavior  to the limited penetration of the perpendicularly polarized electric field into the ultra-narrow nanowires, and similar features are also captured and utilized in meander-type SNSPDs before\cite{berggren_2008_snspd_optical_modeling,simit_2015_polarization}. The absorption rates tend to saturate with further increased nanowire width  (>80\,nm) due to displaced nanowire position from the center of the waveguide, where the evanescent field is the strongest. Based on these results, we choose slightly wider 80\,nm nanowire of 5\,nm-thick in the following discussions, allowing for better electric field penetration into the nanowire, while the nanowire is still narrow and thin enough for obtaining saturated internal efficiencies at the wavelength of 1550\,nm\cite{simit_2017_92p_nbn_detector,delft_2017_92p_nbn_detector,nict_2017_nbtin_snap}.

\subsection{Effect of waveguide parameters}
\label{section:wg_type}

In this subsection, we continue to investigate the absorption rates of the nanowire integrated with waveguides of various types. Following the results of previous sections, we focus on high-index waveguide core material but change the index of upper and bottom claddings, as shown in \cref{fig:perfect_wg_type}(a)-(c), which respectively represent the air-cladding waveguide on SiO\textsubscript{2}, SiO\textsubscript{2}-cladding buried waveguide and fully suspended waveguide. Here, we only consider fundamental TE modes, since higher-order modes provide substantially reduced absorption rates due to larger mode size. Likewise, we do not show the results of partially etched rib waveguides, which have lower absorption rates compared with ridge waveguides because of larger and pulled-down mode profile\cite{pernice_2013_absorption_engineering}. \Cref{fig:perfect_wg_type}(d)-(f) demonstrate simulated NbN nanowire absorption rates versus waveguide width for varying type and thickness. The highest absorption rates are obtained with fairly thin waveguides of 100-140\,nm thickness, which could provide strong evanescent field to couple with the nanowire. In comparison with the other two types of waveguides, the fully suspended waveguide (\cref{fig:perfect_wg_type}(f)) shows significantly stronger absorption for a wide range of waveguide width and thickness due to further confined optical mode with higher index contrast.

\section{H0-type PhC cavity integration}
\label{section:2D}

\begin{figure}[!b]
\centering\includegraphics[width=\linewidth]{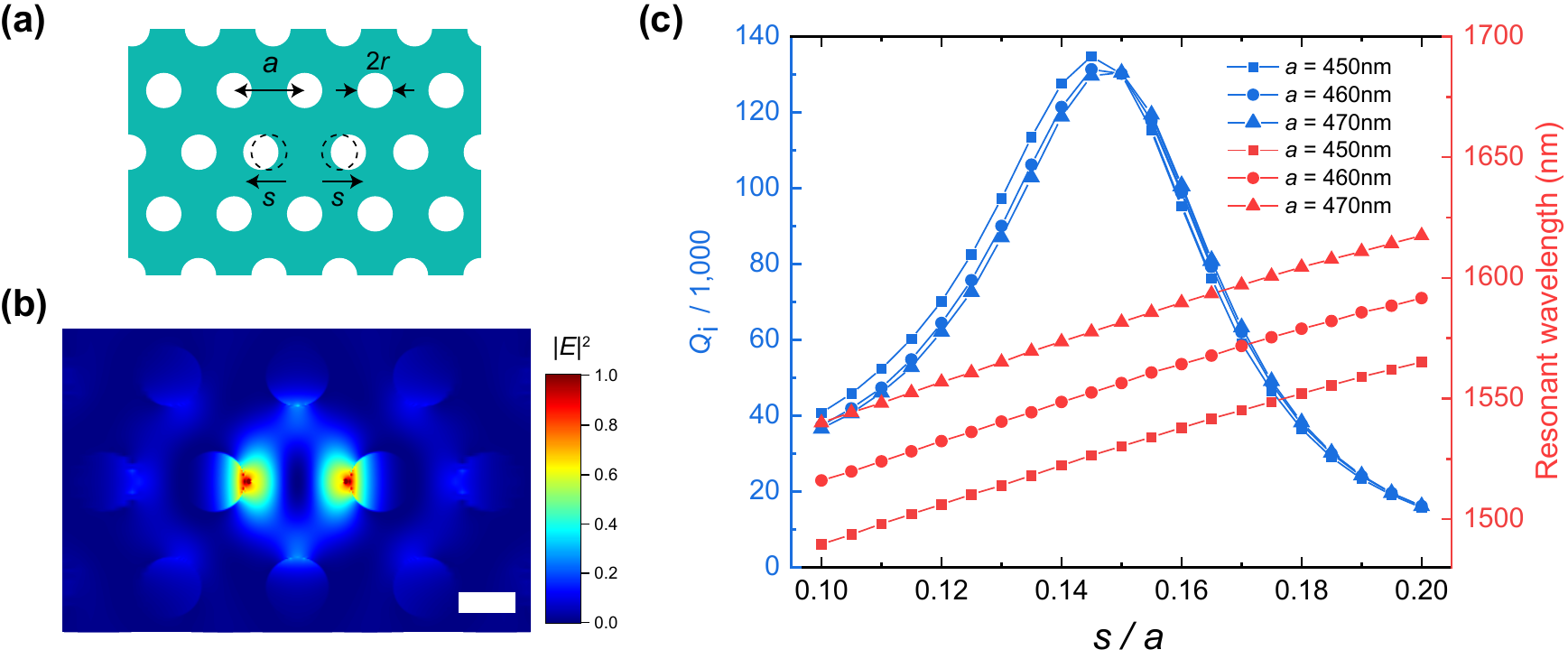}
\caption
{H0-type PhC cavity design and optimization.
(a) Schematic illustration of the H0-type PhC cavity formed by slightly shifting two air holes away from their original positions. $a$, $r$ and $s$ denote the lattice constant, hole radius and the amount of hole shifts, respectively.
(b) Simulated electric field distribution of the H0-type PhC cavity at the resonant wavelength. 
(c) Simulated intrinsic quality factor $Q_\mathrm{i}$ and resonant wavelength of the H0-type PhC cavity versus $s$/$a$ for varying value of $a$. 
}
\label{fig:perfect_H0}
\end{figure}

Among a variety of PhC cavities, we choose H0-type cavity (also referred to as ``zero-cell'' cavity or ``point-shift'' cavity) due to the ultra-small mode volume and ease of integration with nanowire detectors. As shown in \cref{fig:perfect_H0}(a) and (b), the cavity is formed by slightly shifting two air holes away from their original positions by $s$ in the direction of $\Gamma$-K in a two-dimensional hexagonal lattice photonic crystal. Following the previous discussions, we use fully suspended slab made from material of 3.45 index for the PhC cavity design, while the thickness of the slab is chosen to be  220\,nm, which is thicker than the optimal value shown in \cref{fig:perfect_wg_type}(f), considering the trade-off between large enough band gap of the photonic crystal and strong evanescent field at the top surface where the nanowire is situated.

We first start the optimization of the cavity from a lattice constant $a$ of 480\,nm to place the center of the band gap at 1550\,nm wavelength. Then, iteratively tuning $a$, $s$ (shift amount) and $r$ (radii of the air holes), the best intrinsic quality factor $Q_\mathrm{i}$ is obtained with $r=0.25a$. As shown in \cref{fig:perfect_H0}(c), the resonant wavelength increases with $s$ and $a$, while the highest  $Q_\mathrm{i}$ always occurs around $s=0.15a$ independent of $a$. The decoupled $Q_\mathrm{i}$  from $a$ allows us to adjust the resonance of the cavity to the target wavelength by freely tuning $a$. We choose $a=455\,\mathrm{nm}$ for the resonant wavelength of 1550\,nm, and a $Q_\mathrm{i}$ of 132k combined with an ultra-small mode volume of $0.22(\lambda/n)^3$ is achieved by 3D FDTD simulation.

\begin{figure}[!t]
\centering\includegraphics[width=\linewidth]{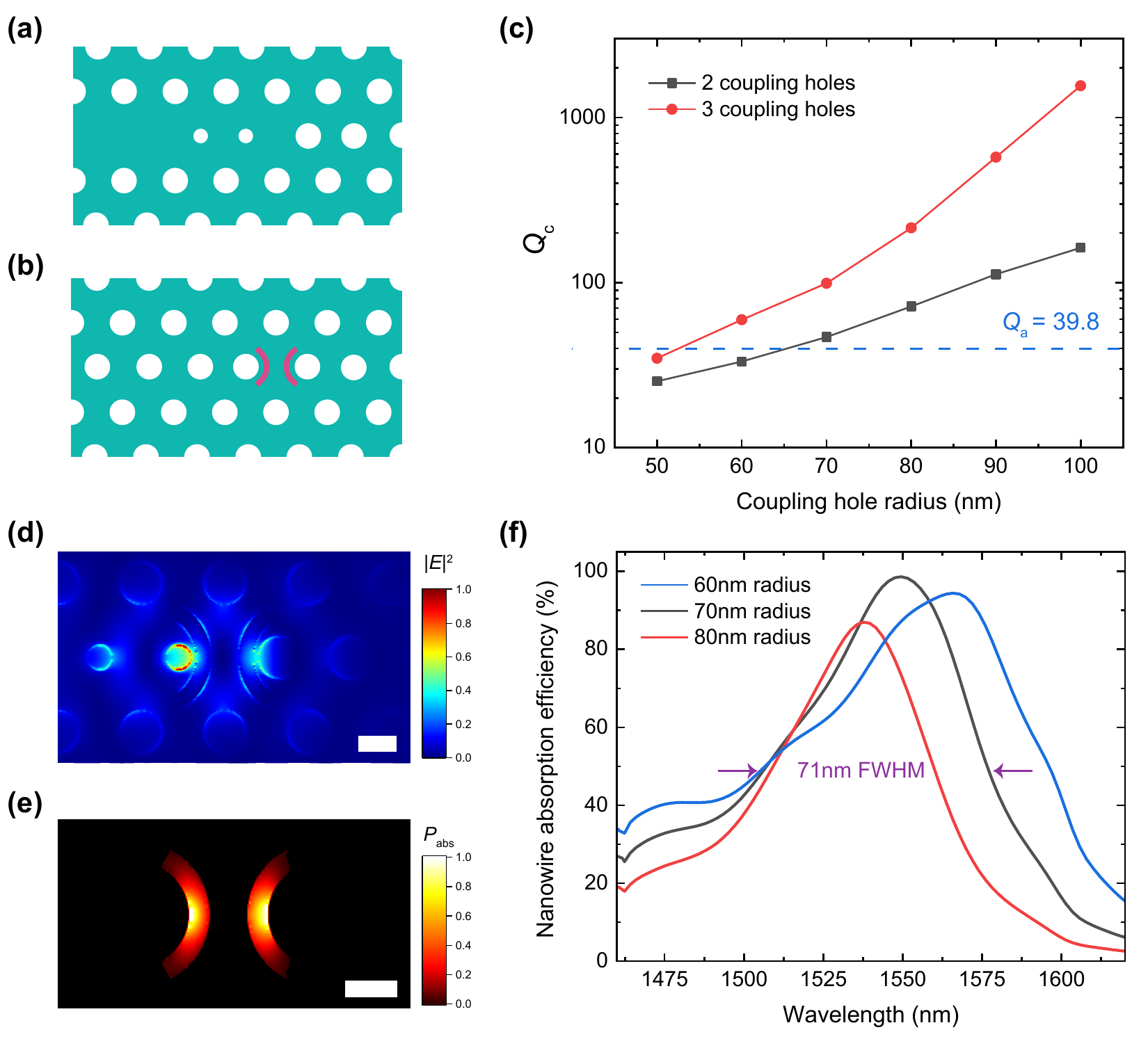}
\caption{
Nanowire absorber integrated with H0-type PhC cavity. 
(a) Schematic of the H0-type PhC cavity with front partial mirror consisting of two smaller air holes.
(b) Schematic of the H0-type PhC cavity with two arc-shaped nanowires embedded inside. 
(c) Simulated coupling quality factor $Q_\mathrm{c}$ of the H0-type PhC cavity as a function of the number and the radius of the coupling holes. The blue dashed line represents the absorption quality factor $Q_\mathrm{a}$ of the cavity with the nanowires loaded. 
(d) Simulated electric field distribution of the critically coupled H0-type PhC cavity at the resonant wavelength.  
(e) Simulated power dissipation density in the nanowires at the resonant wavelength. 
(f) Simulated dependence of the nanowire absorption on the wavelength for varying coupling hole sizes. The FWHM defining the 3\,dB bandwidth of the nanowire detector at the critical coupling condition is marked by a pair of purple arrows. 
All scale bars, 200\,nm.
}
\label{fig:perfect_2D}
\end{figure}

Based on the optimized design of the cavity, we simulate the coupling quality factor ($Q_\mathrm{c}$) and the absorption quality factor ($Q_\mathrm{a}$) by either making a partial reflective mirror for the cavity (\cref{fig:perfect_2D}(a)) or placing a NbN nanowire on the cavity (\cref{fig:perfect_2D}(b)). In \cref{fig:perfect_2D}(a), a W1-type PhC waveguide is formed by removing a row of the holes at the left side of the cavity, and a front partial mirror is inserted in between to couple the waveguide to the cavity. The reflectivity of the mirror or the coupling strength could be tuned by changing the number and the radii of the coupling holes making up the mirror. As shown in \cref{fig:perfect_2D}(b), the NbN nanowire detector comprises two of 120 degree arcs placed atop the slab and adjacent to the inner edges of the two shifted holes forming the cavity. The thickness, width, radius and total length of the nanowire arcs are 5\,nm, 80\,nm, 265\,nm and 1.1\,$\upmu$m, respectively. The closest gap between the nanowire and the hole is set to 30\,nm, considering a typical alignment tolerance in electron-beam exposure process. \Cref{fig:perfect_2D}(c) shows the simulated $Q_\mathrm{a}$ along with $Q_\mathrm{c}$ depending on the number and radii of the coupling holes. With the nanowire loaded as a strong absorber, the cavity shows substantially reduced quality factor ($Q\approx Q_\mathrm{a}=39.8$) in sharp contrast to its $Q_\mathrm{i}$. We could obtain near-unity absorption for the nanowire by engineering $Q_\mathrm{c}$ and fulfilling the critical coupling condition ($Q_\mathrm{c}=Q_\mathrm{a}$, equivalent to $|r|=\alpha$ in \cref{fig:perfect_lossy}(a)). As can be seen in \cref{fig:perfect_2D}(c), the matching condition requires two coupling holes and the optimal hole radius to be approximately 65\,nm.

In order to further investigate the maximum absorption and the frequency response of the nanowire-cavity system, we excite a propagating mode in the coupling PhC waveguide and simulate the absorption by the nanowire as a function of the wavelength around 1550\,nm. As shown in \cref{fig:perfect_2D}(d) and (e), the electric field is strongly confined at the inner edges of the shifted holes, resulting in greatly enhanced power absorption at the center of the nanowire arcs. We note that the electric field and thus the absorption at the ends of the nanowire arcs are very weak, and therefore the nanowires could be quickly tapered to much thicker wires for the electrical connection without introducing much extra absorption in the photon-insensitive area. \Cref{fig:perfect_2D}(f) shows the wavelength-dependent nanowire absorption efficiency normalized to the incident power carried by the coupling waveguide. As the size of coupling holes increases, we could clearly see the transition from over-coupling ($Q_\mathrm{c}<Q_\mathrm{a}$) to under-coupling ($Q_\mathrm{c}>Q_\mathrm{a}$). The critical coupling condition is fulfilled with two 70\,nm-radius holes, and the maximum absorption of 98.6\% is recorded at 1549.6\,nm. We attribute the remaining photon loss to the scattering, which could be further reduced by structure optimization, such as individual tuning of the two coupling hole sizes and positions. From the numerically simulated spectrum, the FWHM or 3\,dB bandwidth at the critical coupling condition is 71\,nm, corresponding to a overall loaded quality factor $Q_\mathrm{l}=21.8$, which is in good agreement to the prediction results by the method of quality factor matching (\cref{fig:perfect_2D}(c)), where $Q_\mathrm{l}=(Q_\mathrm{a}^{-1}+Q_\mathrm{c}^{-1})^{-1} =Q_\mathrm{a}/2=19.9$.  

\section{Discussion and conclusion}
\label{section:comparison}

\renewcommand{\arraystretch}{1.3}
\begin{table*}[!t]
\centering
\caption
{Summary and comparison of SNSPDs with different device structures.} 
\setlength{\arrayrulewidth}{0.3mm}
\begin{tabular}{ |p{1.5cm}||p{4.5cm}|p{1.7cm}|p{1.7cm}|p{2.5cm}| }
 \hline
 Reference& Device type & Nanowire length ($\upmu$m) & 3\,dB bandwidth (nm) & Figure of \newline merit $F$ ($\times\,10^{-3}$)\\
 \hline
 Ref.\cite{simit_2016_broadband_snspd} & Meander nanowire + \newline vertical cavity + metal mirror &1150  & 700  & 0.6  \\
 Ref.\cite{simit_2017_92p_nbn_detector}  & Meander nanowire + \newline vertical cavity + DBR mirror &1350  & 400 & 0.3 \\ 
  Ref.\cite{simit_2019_microfiber_snspd}  & Microfiber-coupled \newline meander nanowire &1100  & 870 & 0.8 \\ 
 Ref. \cite{Bristol_2016_modelling_snspd_wg_cavity} & Racetrack resonator integration &1 & 1&  1 \\
 Ref. \cite{canada_2015_perfect_nanowire_absorber} & 1D-PhC cavity integration & 8.5 & 5.6 & 0.7\\
Ref. \cite{pernice_2016_1D_PhC_detector} & 1D-PhC cavity integration & 1 & 10 & 10 \\
Ref. \cite{pernice_2018_2DPC}& 2D-PhC cavity integration  & 3  & 13.2  & 4.4 \\
 This work & H0-type PhC cavity integration  & 1.1 & 71 & 64.5 \\
 \hline
\end{tabular}
\label{table:perfect_list}
\end{table*}

There is a general trade-off in all cavity-integrated perfect absorber design between the absorption volume and the absorption bandwidth, as derived from \cref{eq:loaded_resonance}. For example, for less absorption volumes (smaller $\alpha$) or shorter nanowires in the case of cavity-integrated SNSPDs, the photons are expected to experience more round-trips inside the cavity to achieve a near-unity absorption, which entails higher finesse of the cavity (larger $Q_\mathrm{l}$ for the fixed cavity mode volume) at the expense of reduced bandwidth. Therefore, we introduce a figure of merit defined as bandwidth-nanowire-length ratio $F = B/NL$ to balance the design compromise between $B$ and $NL$, where $NL$ is the length of the nanowire required for achieving near-unity maximum absorption, and $B$ stands for the 3\,dB bandwidth of the nanowire absorption spectrum. \Cref{table:perfect_list} shows the comparison of our work in terms of $B$, $NL$ and $F$ with other types of recently demonstrated SNSPDs, including meander-type detectors integrated with vertical cavities, detectors integrated with waveguides and various types of on-chip cavities.  In comparison with traditional meander-type SNSPDs,  our H0-type 2D-PhC cavity design shows two orders of magnitude improved $F = 64.5$  with a dramatically reduced nanowire length down to 1.1\,$\upmu$m. In the meanwhile, this design still maintains a significantly enhanced 3\,dB bandwidth of 71\,nm, compared to other cavity-integrated SNSPDs.

It should also be mentioned that the fully suspended HIC waveguide that our designs are based on could be realized either on Si or GaAs platforms, which benefit the realization of photonic circuits in very compact size. In particular, GaAs has drawn great interest of research recently due to its strong $\chi^{(2)}$ and $\chi^{(3)}$ nonlinearity\cite{bower_2018_gaas_on_insulator}, large electro-optic effect\cite{bristol_2014_gaas_quantum_circuit} as well as the capability of on-chip single-photon detector integration\cite{Fiore_2015_GaAs_SNSPD} and  quantum dots growth as on-demand single-photon sources\cite{bristol_2016_gaas_review}. These excellent optical functionalities renders GaAs a very promising candidate platform for realizing fully integrated quantum photonic circuits with the generation, routing, active manipulation and the final detection of single-photons on a single chip\cite{kit_2018_full_hbt,bristol_2016_gaas_review}. 

Combined, we expect that our proposed design will enable a large array of high-performance on-chip single-photon detectors for the future fully integrated quantum photonic circuits, simultaneously offering high efficiency, large bandwidth, ultra-low jitter, ultra-fast gigahertz counting rates as well as high fabrication yield.

\section*{Acknowledgments}
We thank Dr. Xiang Guo and Dr. Linran Fan for the fruitful discussions. We acknowledge funding support from DARPA DETECT program through an ARO grant (No: W911NF-16-2-0151), NSF EFRI grant (EFMA-1640959), AFOSR MURI grant (FA95550-15-1-0029), and the Packard Foundation.

\def\bibsection{\section*{References}}
\bibliographystyle{Risheng}
\bibliography{My_reference}
\end{document}